\documentclass[12pt]{article}
\usepackage{graphicx}
\usepackage{color}
\usepackage{ulem}
\begin{document}
\newcommand{\beq}{\begin{eqnarray}}
\newcommand{\eeq}{\end{eqnarray}}
\newcommand{\ve}{\varepsilon}
\newcommand{\ben}{\begin{enumerate}}
\newcommand{\een}{\end{enumerate}}
\newcommand{\bit}{\begin{itemize}}
\newcommand{\eit}{\end{itemize}}
\newcommand{\vo}{\varepsilon_0}
\newcommand{\vi}{\varepsilon(\infty)}
\newcommand{\bq}{{\bf q}}
\newcommand{\bp}{{\bf p}}
\newcommand{\bk}{{\bf k}}
\newcommand{\bv}{{\bf v}}
\newcommand{\br}{{\bf r}}
\newcommand{\by}{{\bf y}}
\newcommand{\bcr}{{\bf R}}
\newcommand{\bcq}{{\bf Q}}
\newcommand{\bcb}{{\bf B}}
\newcommand{\bce}{{\bf E}}
\newcommand{\bcg}{{\bf G}}
\newcommand{\bcw}{{\bf W}}
\newcommand{\red}[1]{\textcolor{red}{#1}}
\newcommand{\blue}[1]{\textcolor{blue}{#1}}
\title{Theory of Polar Corrections to Donor Binding }
\author{G.D. Mahan$^{\dagger}$ and Kristian Berland$^*$\\
$^\dagger$Department of Physics\\ Pennsylvania State University, University
park, PA\\$^*$Department of Microtechnology and Nanoscience,\\ Chalmers University of
Technology, G\"{o}teborg, Sweden}
\maketitle
\begin{center}
Abstract
\end{center}
We calculate the optical phonon correction to the binding energy of electrons to
donors in cubic materials. Previous theories calculated the Rydberg energy reduced by the
effective mass and the static dielectric
function. They omitted an important energy term from the long-range polarization of
the ionized donor, which vanishes for the neutral donor. They also omitted the donor-phonon interaction. Including these terms yields a new formula for the donor binding energy. 
\newpage
\section{Introduction}
We present a  calculation of the polaron correction to the binding energy $E_D$  of
electrons to donors. This topic was treated originally by
Larsen\cite{larsen,ams,dml2,dml3,dml4,dml5}. We adopt a similar approach, but
include more terms in the theory. 

Prior theories are called the scaled hydrogen model (SHM). In
SHM one takes the binding energy of hydrogen, the Rydberg energy, and scales it
with the effective band mass $m^*$ of the electron, and the dielectric function $\ve$ of the
material. In SI units
\beq
E_D &=& \frac{e^4m^*}{2(4\pi \ve)^2\hbar^2}
\eeq 
For weakly bound donors, the dielectric function is usually taken to be the zero frequency value $\ve(0)$: which
includes the polar contribution from optical phonons.  When the donor binding energy is larger than the optical phonon energy, then some advocate using the high-frequency dielectric constant $\ve(\infty)$. Neither our theory, nor that of
Larsen, were able to derive these expressions for the SHM. We show below that our theory gives the formula for the effective dielectric constant
\beq
\frac{1}{\ve} &=& \frac{5}{16 \vi}+\frac{11}{16\ve(0)} 
\eeq
The fractional factors are unexpected, and are the main result of our calculation. 
The SHM is still in
regular use today\cite{bella,mex,wijn,xia}.

In doing this calculation, we became aware
that prior theories omitted an important energy term: the polarization energy of
the ionized donor. 
An ionized donor is  viewed as a point charge in a polarizable medium.
There is a polarization energy associated with this long-range poential. We
calculate below the exact expression for this energy. When the electron becomes
bound to the donor, the donor becomes neutral, and the long-range polarization
energy vanishes. It is replaced by short-range polarization on the scale of the
Bohr radius of the donor. This difference in polarization energy contributes to
the binding energy of the donor. It is one reason for the fractions in eqn.(2).

We present calculations of donor binding energies for materials with the fcc
(face-centered-cubic) lattice. This includes most III-V and II-VI
semiconductors, and some oxides. For semiconductors with small donor binding energies, we get similar accuracy compared to the SHM. For the oxides, in particular, our theory gives
much larger binding energies than does the SHM, but neither agree with experiment.   Our
theory may also apply to the binding of holes to acceptors.  However, the
degeneracy of most valence bands makes this case more complicated, and we do not
consider  the acceptor case.  

Landau and Pekar\cite{landau1,pekar1,lp1} developed the first theory of polarons for a free electron. They did a variational calculation on an electron eigenfunction $\phi(\br)$. They calculated the phonon displacement $\delta Q$ in response to the  charge density of this eigenfunction, and then calculated the polaron energy $\propto (\delta Q)^2$. This phonon excess energy then became part of the variational energy determining the electron eigenfunction. The Landau-Pekar theory later became recognized as the strong-coupling theory of polarons\cite{mpp1}. Here we adopt the same procedure to calculate the response of the phonons to the electron bound to the donor. Instead of the Gaussian eigenfunction used by Landau and Pekar, we use the exponential eigenfunction typical of a donor. Otherwise our calculation follows the same procedures. Obviously, our theory is also a strong coupling theory, which applies when $\vi$ and $\ve(0)$ are very different. 

\section{Theory}
We employ an approach based on our earlier work\cite{gdm1,gdm2}. 
The Hamiltonian  includes optical phonons,  donors at $\bcr_l$ of charge $q>0$,
and an electron of charge $e<0$, effective mass $m^*$, which may be bound to a
donor. The optical phonons have a displacement $\bcq_j$, a Szigeti charge $e^*$,
which creates a dipole $e^*\bcq_j$. 
\beq
H &=& \frac{p^2}{2m} +\sum_l\frac{eq}{4\pi \vi |\br-\bcr_l|}+\frac{q^2}{4\pi
\vi}\sum_{l>l'}\frac{1}{|\bcr_l-\bcr_{l'}|}+\sum_j\left[\frac{P_j^2}{2M}+\frac{K
}{2}Q_j^2\right]\nonumber\\&& + V_{pd} +V_{ep}+ V_{pp}\label{eqn1}
\eeq
where:
\bit
\item The interaction between the phonons and the donors is
\beq
V_{pd} &=& -\frac{e^*q}{4\pi\vi}\sum_{j,l}
\frac{\bcq_j\cdot(\bcr_j-\bcr_l)}{|\bcr_j-\bcr_l|^3}
\eeq
\item The interaction between the electrons and the optical phonons is 
\beq
V_{ep} &=& -\frac{ee^*}{4\pi\vi}\sum_{j,l}\int d^3rn(r)
\frac{\bcq_j\cdot(\bcr_j-\bcr_l-\br)}{|\bcr_j-\bcr_l-\br|^3}
\eeq
where $n(r)$ is the charge density of an electron  bound to the donor. This reduces to the usual Fr\"{o}hlich
interaction for electrons in a free energy band\cite{mpp3}. 
\item The dipole-dipole interaction between the phonons is
\beq
V_{pp} &=& -\frac{(e^*)^2}{8\pi \vi}\sum_{i\ne
j}\bcq_i\cdot\phi_{ij}\cdot\bcq_j\\
\phi_{ij} &=&
\phi_{\mu\nu}(R_{ij})=\frac{\delta_{\mu\nu}}{R_{ij}^3}-\frac{3R_{ij,\mu}R_{ij,
\nu}}{R_{ij}^5}
\eeq
The lattice transform of this interaction is \cite{gdm1}
\beq
T_{\mu\nu}(\bk) &=& \sum_{j\ne 0} e^{i\bk\cdot\bcr_j}\phi_{\mu\nu}(\bcr_j)\\
\lim_{ka<<1} T_{\mu\nu}(\bk) &=&
-\frac{4\pi}{\Omega_0}\left[\frac{k_{\mu}k_{\nu}}{k^2}-\frac{\delta_{\mu\nu}}{3}
\right]
\eeq
where $\Omega_0$ is the volume of the unit cell.
We will usually take the limit of small wave vector in these expressions.
\eit

\section{Charged Donor}
First  we evaluate the energy of an isolated donor, without an electron nearby. 
We transform the vibrations to collective coordinates
\beq
\bcq_j &=& \frac{1}{\sqrt{N}}\sum_{\bk}e^{i\bcr_j\cdot\bk}\bcq_{\bk}
\eeq
The potential energy terms for LO phonons are 
\beq
V &=& \sum_{\bk}\left[\frac{K_L}{2}[\hat{k}\cdot \bcq_k]^2 + \beta_k\hat{k}\cdot
\bcq_k\right]\label{eqn11}\\
K_L &=& K + \frac{2}{3}\frac{(e^*)^2}{\Omega_0\vi}
\eeq
 The second term in $K_L$ comes from the dipole-dipole interaction. 
The lattice transform of $V_{pd}$ is 
\beq
V_{pd} &=& -\frac{qe^*}{4\pi
\vi\sqrt{N}}\sum_{\bk}\bcq_k\cdot\bcw_{\bk}\sum_{l}e^{i\bk\cdot\bcr_l}\\
\bcw_{\bk} &=& \sum_{j\ne 0} e^{i\bk\cdot\bcr_j}\frac{\bcr_j}{R_j^3}\approx
\frac{4\pi i}{\Omega_o}\frac{\bk}{k^2}
\eeq
where the latter limit is for long wavelength. In this case the expression for
$\beta_{\bk}$ is
\beq
\beta_k &=& -\frac{qe^*i}{\vi k\Omega_0\sqrt{N}}\sum_{l}e^{i\bk\cdot\bcr_l}
\eeq

The phonon displacements enter as $(\hat{k}\cdot\bcq_k)$, where the notation emphasizes that only
longitudinal optical phonons are involved. The transverse $(\omega_{TO})$ and
longitudinal $(\omega_{LO})$  optical phonon frequencies are defined as
\beq
\omega_{TO}^2 &\equiv& \frac{K_T}{M}= \frac{K}{M}-\frac{1}{3}\omega_i^2\\
\omega_{LO}^2 &=& \frac{K_L}{M} =\frac{K}{M} + \frac{2}{3}\omega_i^2, \;\;
\omega_i^2=\frac{(e^*)^2}{M\Omega_0\vi}
\eeq
The static dielectric function is
\beq
\ve(0) &=& \vi\left[1 + \frac{\omega_i^2}{\omega^2_{TO}}\right]\label{eqn12}
\eeq

We complete the square on the interaction in eqn.(\ref{eqn11}), 
\beq
V &=& \frac{K_L}{2}\sum_{\bk}\left[\hat{k}\cdot\bcq_k+
\frac{\beta_k}{K_L}\right]^2-\frac{1}{2K_L}\sum_{\bk} |\beta_{k}|^2
\eeq 
The last term is the effective interaction between the optical phonons and the
donor, due to the phonon polarization
\beq
V' &=& -\frac{1}{2K_L}\sum_{\bk} |\beta_{k}|^2\\
&=& -\frac{q^2(e^*)^2}{2K_L\vi^2\Omega_0}\int
\frac{d^3k}{(2\pi)^3}\frac{1}{k^2}\sum_{ll'}e^{i\bk\cdot(\bcr_l-\bcr_{l'})}
\eeq
For $\bcr_l\ne\bcr_{l'}$ the wave vector integral gives the effective
interaction between the two donors
\beq
V_{ll'}' &=& -\frac{q^2}{4\pi \vi
}\frac{\omega_i^2}{\omega^2_{LO}}\frac{1}{|\bcr_l-\bcr_{l'}|}
\eeq
We combine this term with the direct interaction in eqn.(\ref{eqn1}) to find
\beq
V_{ll'} &=& \frac{q^2}{4\pi \vi
}\left[1-\frac{\omega_i^2}{\omega^2_{LO}}\right]\frac{1}{|\bcr_l-\bcr_{l'}|}\\
&=& \frac{q^2}{4\pi \ve(0) }\frac{1}{|\bcr_l-\bcr_{l'}|}\\
\frac{1}{\ve(0)} &=& \frac{1}{ \vi
}\left[1-\frac{\omega_i^2}{\omega^2_{LO}}\right]
\eeq
The latter identity follows directly from eqn.(\ref{eqn12}). The effective
Coulomb interaction between ionized donors is screened by the static dielectric
function. 

The case that $l=l'$ gives the energy of a single donor from the phonon
polarization. It has the approximate value of 
\beq
\int \frac{d^3k}{(2\pi)^3}\frac{1}{k^2} &=&
\frac{4\pi}{(2\pi)^3}\frac{\pi}{\tilde{a}}=\frac{1}{2\pi \tilde{a}}\\
V' &=& -\frac{q^2(e^*)^2}{4\pi \tilde{a}K_L\vi^2\Omega_0}\label{eqn25}
\eeq
where $\tilde{a}$ is approximately a lattice constant. An exact expression for $\tilde{a}$
 is given below. This energy term comes from the polarization induced by a single charged
donor, which reduces its energy. 

\section{Electron Bound to Donor}
We repeat the above calculation but include  an electron bound to a donor at $\bcr_l$. 
We do this calculation of the donor binding energy using a variational parameter
$\alpha$ defined as
\beq
\psi(r) &=& \sqrt{\frac{\alpha^3}{\pi a_0^3}}\exp[-\alpha r/a_0]\\
n(r) &=& |\psi(r)|^2
\eeq
where the Bohr radius is given by 
\beq
a_0 &=& 4\pi \vi\frac{\hbar^2}{e^2m^*}
\eeq
 and $m^*$ is the band effective mass of the conduction electron.
Larsen\cite{larsen} did a similar calculation but omitted the interaction
$V_{pd}$ which is the term that polarizes the lattice in the absence of an electron. The inclusion of this important term is crucial for including all essential interactions. 
Again we transform the vibrations to collective coordinates
\beq
\bcq_j &=& \frac{1}{\sqrt{N}}\sum_{\bk}e^{i\bcr_j\cdot\bk}\bcq_{\bk}
\eeq
The potential energy terms  in the limit of long wave-length are 
\beq
V &=& -\sum_{\bk}\left[\frac{K_L}{2}[\hat{k}\cdot Q_k]^2 + \gamma_k\hat{k}\cdot
\bcq_k\right]\\
\gamma_k &=& \frac{ie^*}{\vi k\Omega_0\sqrt{N}}\left[q +
e\Lambda(ka_0/2\alpha)\right]\sum_{l}e^{i\bk\cdot\bcr_l}\\
\Lambda(x) &=& \frac{1}{(1+x^2)^2}
\eeq
The first term in $\gamma_k$ comes from the donor-phonon interaction, and the
second comes from the electron-phonon interaction. The donor-phonon interaction
is evaluated in the long wavelength limit. We complete the square on the
interaction, and also set $q=-e$
\beq
V &=& \frac{K_L}{2}\sum_{\bk}\left[\hat{k}\cdot\bcq_k+
\frac{\gamma_k}{K_L}\right]^2-\frac{1}{2K_L}\sum_{\bk} |\gamma_{k}|^2
\eeq 
The last term is part of the effective interaction between the electron and the
donor, due to the phonon polarization
\beq
V" &=& -\frac{1}{2K_L}\sum_{\bk} |\gamma_{k}|^2\\
&=& -\frac{e^2(e^*)^2}{2K_L\vi^2\Omega_0}\int
\frac{d^3k}{(2\pi)^3}\frac{1}{k^2}\left[1-\Lambda(ka_0/2\alpha)\right]^2\sum_{
ll'}e^{i\bk\cdot(\bcr_l-\bcr_{l'})}\label{eqn35}
\eeq
We evaluate this expression below in several limits. There are two types of
terms. If $l\ne l'$ then it is the phonon-induced interaction between two
neutral donors. If $l=l'$ it is the polaron correction to the binding energy of
a single donor. Note that $\Lambda(x=0)=1$ so the integrand vanishes at $k=0$.
There is no  long-range Coulomb interaction between neutral donors.

\subsection{Interactions between donors} 
First consider the case that $l\ne l'$ so the expression gives the
phonon-induced interaction energy between two neutral donors.
For the effective mass limit, where the orbit radius $a_0$ covers many lattice constants,
we can extend the limit of integration to infinity. Change to dimensionless
variables $x=ka_0/(2\alpha), y = 2\alpha R_{ll'}/a_0$, and we get
\beq
V_{pp'}" &=& -\frac{e^2}{4\pi\vi a_0}\frac{\omega_i^2}{\Omega^2_{LO}}I(y)\\
I(y) &=& \frac{2}{\pi y} \int_0^{\infty}dx \sin(xy)
\frac{x^3(2+x^2)^2}{(1+x^2)^4}
\eeq
The above integral can be evaluated by contour integration. It gives a term 
\beq
I(y) &=& \frac{e^{-y}}{y}\sum_{m=0}^3a_my^m
\eeq
The interaction decays exponentially, which is typical of the interaction
between neutral charge distributions, when neglecting correlations. 

\subsection{Neutral Donor \#1}
We evaluate eqn.(\ref{eqn35}) for the case that $l=l'$. Expand the factor
\beq
(1-\Lambda)^2 &=& 1-2\Lambda+\Lambda^2
\eeq
The first term ("1") we evaluate as eqn.(\ref{eqn25}), so the polarization
energy of the donor is unchanged. The other two terms contribute to the donor binding energy.  In the effective mass limit, the last two
terms are evaluated assuming that the $k$-integral goes to infinity. 
\beq
\frac{4\pi}{(2\pi)^3}\int_0^{\infty}dk[-2\Lambda + \Lambda^2] &=&
-\frac{\alpha}{2\pi a_0}\left[1-\frac{5}{16}\right]= -\frac{11\alpha}{32\pi a_0}
\eeq
The variational energy for the donor binding is
\beq
{\cal E}(\alpha)&=& E_{Ry}[\alpha^2 -2\alpha(1-\lambda)],\;\;\; \lambda
=\frac{11}{16}\frac{\omega_i^2}{\omega^2_{LO}}\label{eqn41}
\eeq
Minimizing with respect to $\alpha$ gives
\beq
\alpha_0 &=& 1-\lambda = \frac{5}{16}+\frac{11}{16}\frac{\vi}{\ve(0)}\\
{\cal E}(\alpha_0) &=& -E_{Ry}(1-\lambda)^2
\eeq
For weakly bound donors, the dielectric screening is not given by either
$\ve_{\infty}$ or $\ve(0)$. Instead, it is 5/16 of $\ve_{\infty}$ and 11/16 of
$\ve(0)$. 

An interesting result is obtained if we neglect $\Lambda^2$. In that case 
the total variational energy for the electron bound to the donor is
\beq
{\cal E}(\alpha) &=& \frac{\hbar^2\alpha^2}{2m^*a_0^2} -\frac{e^2\alpha}{4\pi\vi
a_0}(1 - \frac{\omega_i^2}{\omega_{LO}^2})\\
&=& \frac{\hbar^2\alpha^2}{2m^*a_0^2} -\frac{e^2\alpha}{4\pi\ve(0) a_0}
\eeq

Varying $\alpha$ gives the minimum 
\beq
\alpha_0 &=& \left(\frac{\vi}{\ve(0)}\right)\\
{\cal E}(\alpha_0) &=& -E_{Ry}\left(\frac{\vi}{\ve(0)}\right)^2
\eeq
The donor binding energy is screened by the zero frequency dielectric function,
which includes the contribution from the optical phonons. This is the usual
form of the SHM, which is obtained only by neglecting $\Lambda^2$. Since neglecting
$\Lambda^2$ is a poor approximation, the SHM is not accurate. 

\subsection{Neutral Donor \#2}
Here we evaluate the screening corrections to the neutral donor by a different
method, which calculates all energy terms in real space. This has the advantage of not having to make a long-wavelenth approximation. The final expression should be more accurate. 

The interaction $V_{ep}$ is rewritten as 
\beq
V_{ep} &=& \frac{ee^*}{4\pi \vi
\sqrt{N}}\sum_{\bk}\left(\sum_{l}e^{i\bk\cdot\bcr_l}\right)\sum_{j\ne
0}e^{i\bk\cdot\bcr_j}\bcq_{\bk}\cdot\vec{\nabla}_{R_j} f(R_j)\\
f(R) &=& \int d^3r \frac{n(r)}{|\bcr-\br|}=\frac{1}{R}-\frac{1}{R}e^{-2\alpha
R/a_0}\left(1 + \frac{\alpha R}{a_0} \right)
\eeq
When $q=-e$ the $1/R$ term cancels $V_{pd}$. The combined terms give for
$\gamma_{\bk}$
\beq
\vec{\gamma}_{\bk} &=& -\frac{ee^*}{4\pi \vi
\sqrt{N}}\left(\sum_{l}e^{i\bk\cdot\bcr_l}\right)\sum_{j\ne
0}e^{i\bk\cdot\bcr_j}\vec{\nabla}_R\left\{\frac{e^{-2\alpha R/a_0}}{R}(1+\alpha
R/a_0)\right\}_{R=R_j}\nonumber\\
&=&  \frac{ee^*}{4\pi \vi
\sqrt{N}}\left(\sum_{l}e^{i\bk\cdot\bcr_l}\right)\sum_{j\ne
0}e^{i\bk\cdot\bcr_j}\hat{R}_j\frac{e^{-2\alpha R_j/a_0}}{R_j^2}(1+2\alpha
R_j/a_0 +2\alpha^2R_j^2/a_0^2)\nonumber
\eeq
The next step is to take $\sum_{\bk}|\gamma_{\bk}|^2$. There is a double sum
over $(j,j'$). When $l=l'$  the summation over $\bk$ forces $\bcr_j=\bcr_{j'}$.
The final answer for $l=l'$ is
\beq
V"_{pp}&=& -\frac{e^2(e^*)^2}{2K_L(4\pi \vi)^2}\sum_{j\ne 0}\frac{e^{-4\alpha
R_j/a_0}}{R_j^4} (1+2\alpha R_j/a_0 +2\alpha^2R_j^2/a_0^2)^2
\eeq
This answer is only slightly different than case \#1. The summation over
$\bcr_j$ converges to a finite value. There is no ionized donor energy as in
eqn.(\ref{eqn25}). That is sensible, since the neutral donor should not attract
long-range polarization. We rewrite the above expression using the nearest neighbor distance $d$ as a unit of length
\beq
V"_{pp}&=&-E_{Ry}\frac{\omega_i^2}{\omega_{LO}^2}\frac{a_0}{d}F(\alpha
d/a_0)\label{eqn51}\\
F(\alpha d/a_0) &=& \frac{\Omega_0 d}{4\pi}\sum_{j\ne 0}\frac{e^{-4\alpha
R_j/a_0}}{R_j^4} (1+2\alpha R_j/a_0 +2\alpha^2R_j^2/a_0^2)^2\label{F(x)}
\eeq
An important result is the screening energy of the ionized donor, which is found
by setting $a_0=\infty$
\beq
V' &=& -\frac{e^2}{8\pi \vi d}\left(\frac{(e^*)^2}{\vi K_L\Omega_0}\right)
F(0)\label{vpp}
\eeq
This formula is exact, and is an improvement to eqn.(\ref{eqn25}). For the fcc
lattice $F(0)=1.425780$ using Ewald methods, which are described in the
Appendix. By comparing the two expressions, we determine that 
\beq
\tilde{a} &=& a\frac{\sqrt{2}}{F(0)}=0.992 a
\eeq
 The energy of the ionized donor must be subtracted from the result of equation
(\ref{eqn51}), which gives the final donor binding energy
\beq
{\cal E}(\alpha) &=& E_{Ry}\{\alpha^2 -2\alpha -\eta [F(\alpha
d/a_0)-F(0)]\}\label{eqn56}\\
\eta &=& \frac{(e^*)^2a_0}{ \vi K_Ld\Omega_0}=\frac{a_0}{
d}\frac{\omega_i^2}{\omega_{LO}^2}
\eeq
For a given crystal structure,
$F(x)$ is a function of $x=\alpha d/a_0$. The self-consistent variational
equation for the coupling constant $\alpha_0$ at minimum energy is
\beq
\alpha_0 &=& 1 + \frac{1}{2}\frac{\omega_i^2}{\omega_{LO}^2}F'(x)_{x=\alpha_0
d/a_0}\label{alfeqn}
\eeq
where $F'(x)=dF/dx<0$. 

For many semiconductors the value of $x$ is small, and it is adequate to take
the limit of small $x$
\beq
F(x)-F(0) &=& x F'(0) + O(x^3)
\eeq 
Using Ewald methods we show for the fcc lattice that $F'(0)=-11/8$, which makes
eqn.(\ref{eqn56}) identical to eqn.(\ref{eqn41}). The derivation in the appendix suggests that this result $(F'(0)=-11/8)$ is valid for all lattices.   For weakly bound donors our
two derivations give the same result. Equations(\ref{eqn56}) and (\ref{alfeqn})
are useful for materials with larger values of $x$.

\subsection{Atomic Limit for Donor}
One case is when the donor is tightly bound, so that $d/a_0>1$. Even then the  term
$F(x)$ is not small  and cannot be neglected.  For example, $F(1)=0.35$.  

\subsection{Ferroelectrics}
Many ferroelectrics have the feature that as one nears the transition
temperature $T_c$ then $\ve(0)$ diverges while $\vi$ remains  constant.
This is the result of $\omega_{TO}$ going to zero, while $\omega_{LO}$ remains
 constant. In our first model, as $\omega_{TO}\rightarrow 0$ then
$\omega_{LO}\rightarrow \omega_i$. The donor binding energy does  not vanish,
but is given by $E_{Ry}(5/16)^2 \approx 0.1 E_{Ry}$.  

\section{fcc Lattice}
We present some calculations for the fcc lattice, where $d=a/\sqrt{2}$ and
$\Omega_0=a^3/4=d^3/\sqrt{2}$. We discuss the evaluation of the function $F(x)$,
in eqn.(\ref{F(x)}), and its derivative $F'(x)$. 
\beq
F(x)  &=& \frac{\Omega_0 d}{4\pi}\sum_{j\ne 0}\frac{e^{-4x R_j/d}}{R_j^4}
[1+2xR_j/d +2x^2(R_j/d)^2]^2
\eeq
Define the summation variable as $y_j=R_j/d$ which starts out as one for the
twelve nearest neighbors. 
\beq
F(x)  &=& \frac{1}{4\pi\sqrt{2}}\sum_{j\ne 0}\frac{e^{-4x y_j}}{y_j^4} (1+2xy_j
+2x^2y_j^2)^2\label{56}
\eeq
One way to evaluate eqn.(\ref{56}) is to sum all neighbors out to a distance
$y_x$. However, the expression converges very slowly for small values of $x$. 
We performed the computer summations, out to a distance $y_x$, and  noticed that
they seem to scale  as $-1/y_x$. We did a least squares fit to the expression
$F(0) = C-B/y_x$, which gave $C = 1.425781$ which is very close to the Ewald
result. Using Ewald methods in the Appendix we obtain the same value.

A similar process determines $F'(x)$
\beq
F'(x) &=& -\frac{8x^2}{4\pi\sqrt{2}}\sum_{j\ne 0}\frac{e^{-4x y_j}}{y_j}(1
+2xy_j + 2x^2y_j^2)\label{eqn62}
\eeq
Using Ewald methods we found the exact expression $F'(0)=-11/8$. In the appendix we derive the next terms in the power series in $x$
\beq
F(x) &=& F(0) -\frac{11}{8}x + \frac{B}{3} x^3 +O(x^4)
\eeq where $B= 1.4595 $. 
With these values we constructed the graph of $F(x)$  shown in
figure 1. 
\begin{figure}
\centering
\includegraphics[width=8cm]{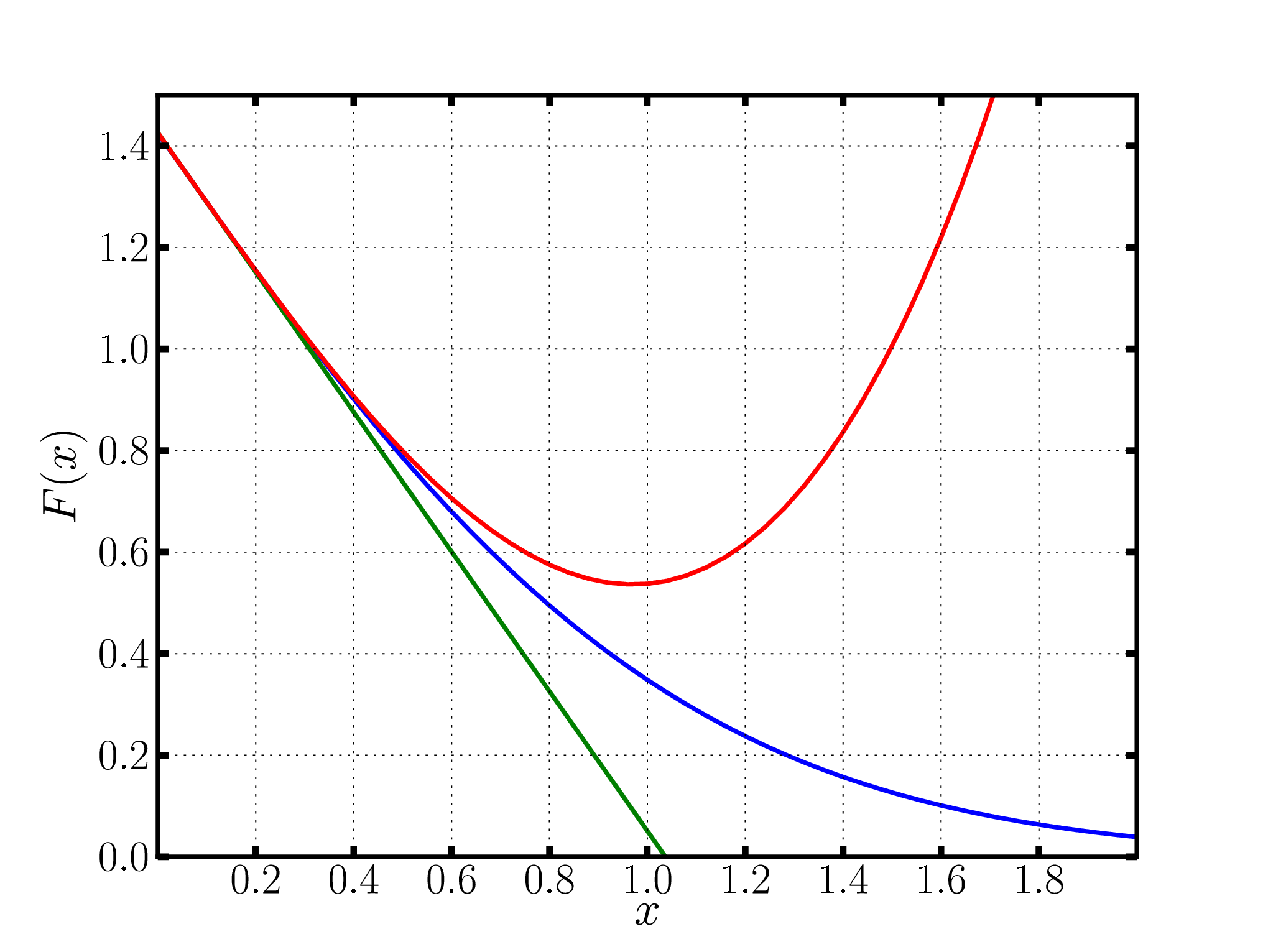}
\caption{$F(x)$ for the fcc lattice. The lower dashed curve shows the two first terms of the expansion in $x$, while the upper shows  the results of the first four terms in $x$. }
\end{figure}

The eigenvalue equation (\ref{alfeqn}) involves the function $F'(x)$, which we have evaluated from equation (\ref{eqn62}). For fitting purposes, it is useful to have an analytic function that approximates these numberical values. The following expression is accurate to $O(0.3\%)$
\beq
F'(x) &\approx& -\frac{24 x^2}{\pi \sqrt{2}}\left[ e^{-4x}(1+2x+2x^2)-e^{-1.3 \xi x}\right]+Bx^2 e^{-\xi x}\nonumber\\
&& -\frac{11}{8}e^{-\xi x}\left[1+\xi x+\frac{1}{2}(\xi x)^2\right]
\eeq
where $B$ is defined above, and the fitting parameter is $\xi=5.09490$. The first term on the right is the asymptotic limit, from the nearest neighbor contribution in eqn.(\ref{eqn62}). 

We have used our theory to calculate the donor binding energies in several
crystals, as shown in table 1. Data is from Landolt-B\"{o}rnstein\cite{lb}. Only
cubic crystals are included, and only those whose conduction band minimum is at
the center of the Brillouin zone, so the effective mass is isotropic. For the
oxides, the binding is sufficiently large that we solved eqn.(\ref{alfeqn})
self-consistently. This gives a much larger binding energy compared to the SHM.
\begin{table}
\begin{tabular}{c|lrrrrrc}
Crystal & $m^*$ & $\vi$ & $\ve(0)$ & $E_{Ry}$ & $E_D$ (SHM) & $E_D$ (Th'y)&
$E_D$(Exp)\\  \hline
GaAs & 0.0665 & 10.9 & 12.5 & 7.6 &   5.8  &6.3 & 5.8-5.9\\
GaSb & 0.0396 & 14.4 & 15.7 & 2.6 &   2.3    &2.4  & \\
InP & 0.079 & 9.6 & 12.6 & 11.7 &     6.8   &8.2 & 7.1 \\
InAs & 0.023 & 12.25 & 15.15 & 2.1 & 1.4	& 1.6& \\
InSb & 0.0145 & 15.68 & 17.5 & 0.80 &   0.60  &0.69 & 0.7\\
ZnS & 0.27 & 5.2 & 8.9 & 136 & 	46	&69 & 30 \\
ZnSe & 0.16 & 6.0 & 8.8 & 60.4 & 28.1	&36.9 & 25-29\\
ZnTe & 0.12 & 7.3 & 10.0 & 30.6 & 16.3	&20.3 & 18.3\\
CdTe & 0.09 & 7.4 & 10.6 & 22.3 & 10.9	&14.0 & 22.0 \\
MgO & 0.35 & 3.0 & 9.9  & 313 & 48.6 & 113 &   \\
CaO & 0.50 & 3.1 & 12.1 & 708 & 69	&248 & 3100\\
SrO & 0.54 & 3.5 & 16.2 & 600 & 191	&401 & 2600\\
BaO & 0.59 & 3.56 & 37.4 & 633 & 147	&394 & 2000 
\end{tabular}
\caption{Donor binding energies in meV. Only cubic crystals are included. Data
from \cite{lb}.}
\end{table}
Both our theory, and the SHM, are poor for the oxides, since electrons are too tightly bound for effective mass theory to be valid. We show these results only to demonstrate that for large values of $x$ our theory is different than SHM.

As stated in the introduction, we regard our theory as a strong-coupling result, which is valid when the two dielectric functions $[\vi$ and $\ve(0)]$ are very different. This is not the case for most III-V semiconductors, and our theory does not do well for these materials. However, we expect our theory to apply quite well to the II-VI semiconductors, which are generally more polar than the III-Vs. Table 1 shows our theory does relatively well for these materials. 
\section{Discussion}
We have evaluated the polaron corrections to the binding energy of electrons to
donors. We include the following interactions: electron-phonon, donor-phonon,
and phonon-phonon. Two variational calculations gave identical results in the
limit of a small binding energy. For the case of a large binding energy, one
method is exact, but gives an equation for the variational parameter that must
be solved self-consistently. 

Most textbooks give that the donor binding energy as
\beq
E_D &=& -\frac{e^4 m}{2 [4\pi \ve(0)]^2 \hbar^2}
\eeq
where $m$ is the band effective mass of the electron, and $\ve(0)$ is the static
dielectric function. We get a different expression
\beq
E_D &=& -\frac{e^4 m}{2 [4\pi \tilde{\ve}]^2 \hbar^2}\\
\frac{1}{\tilde{\ve}} &=& \frac{5}{16 \vi}+\frac{11}{16
\ve(0)} 
\eeq 
which includes both dielectric constants $[\vi, \ve(0)]$. The effective Bohr radius is $\tilde{a}_0 = 4\pi \tilde{\ve}\;\hbar^2/me^2$. We expect our theory applies in the strong-coupling limit, when $\vi$ and $\ve(0)$ are very different. 

There is also the
question of whether the effective mass $m^*$ includes polaron corrections, or is
just the bare band mass. In our theory it is the bare band mass. The above
equation applies only to the case of $E_D<\hbar\omega_{LO}$. For materials with
larger binding energies, one must solve a nonlinear equation to determine the
parameters of the binding energy. We used our theory to calculate the donor
binding energy of several materials with the fcc crystal structure.

One might consider a similar calculation for the electronic polaron effects: the
terms that make $\vi$ differ from $\ve_0$. Here one would start with a bare
interaction, screened by the vacuum dielectric constant $\ve_0$, and consider
how the electronic screening changes the donor binding energy. That is a
different calculation than we have done here, since the electronic screening is
in different places in the crystal. For the oxides, and other ionic crystals,
the electronic polarization resides mostly with the anions\cite{ms}. However,
for covalent materials, it resides in the bonds between ions. 

One of us (GDM) thanks the Erasmus Mundus program of the European Union for
funding his trip to Sweden, where this project was initiated. He also thanks 
professors Per Hyldgaard and Elsebeth Schr\"{o}der for hosting his visit to
Chalmers University of Technology  in G\"{o}teborg. 

\section*{Appendix: Ewald Summation}
The functions $F(x), F'(x)$ are evaluated using an Ewald method\cite{CMiN}.
We start with an evaluation of $F(0)$ since it is easy.
\beq
F(0) &=& \frac{\Omega_0d}{4\pi}\sum_{j\ne
0}\frac{1}{R_j^4}=2\frac{\Omega_0d}{4\pi}\sum_{j\ne 0}\int_0^{\infty} dt t^3
e^{-t^2R_j^2}\\
&=& F_L +F_R\\
F_L &=& 2\frac{\Omega_0d}{4\pi}\sum_{j\ne 0}\int_{\eta}^{\infty} dt t^3
e^{-t^2R_j^2} = \frac{\Omega_0d}{4\pi}\sum_{j\ne 0}\frac{1}{R_j^4}e^{-(\eta
R_j)^2}[1+(\eta R_j)^2]\\
F_R &=& 2\frac{\Omega_0d}{4\pi}\sum_{j\ne 0}\int_0^{\eta} dt t^3 e^{-t^2R_j^2}
\eeq
where $\eta=C/d$. 
The  term $F_L$  we leave as is, since it converges rapidly in real space. The 
term $F_R$ is changed to a summation over reciprocal lattice vectors by creating
a periodic function of position $\br$ 
\beq
V_R(\br) &=& 2\sum_{j}\int_0^{\eta} dt t^3
e^{-t^2(\bcr_j-\br)^2}=\sum_{\bcg}v(\bcg)e^{i\bcg\cdot\br}\\
v(\bcg) &=& \frac{2}{\Omega_0}\int_0^{\eta}t^3dt\int d^3r
e^{-i\bcg\cdot\br}e^{-t^2r^2}\\
&=& \frac{2\pi^{3/2}}{\Omega_0}\int_0^{\eta}dt
e^{-G^2/4t^2}=\frac{G\pi^{3/2}}{\Omega_0}\int_0^{2\eta/G}ds e^{-1/s^2}\\
F_R &=&
\frac{\Omega_0d}{4\pi}\left[\frac{2\pi^{3/2}\eta}{\Omega_0}+\sum_{\bcg\ne
0}v(\bcg)-\frac{\eta^4}{2}\right]
\eeq
The first term in $F_R$ is the $v(0)$, and the last term is from subtracting the
$\bcr_j=0$ term from $V_R(0)$. 
\begin{table}
\begin{tabular}{r|rr|r}
C & $F_L$ & $F_R$ & $F$ \\ \hline
0.7 & 0.812176 & 0.613604 & 1.425780\\
0.9 & 0.646635 & 0.779145 & 1.425780\\
1.1 & 0.492121 & 0.933659 & 1.425780\\
1.3 & 0.353968 & 1.071812 & 1.425780\\
\end{tabular}
\caption{Ewald summation for $F(0)$. }
\end{table}
Table 2 shows the separate contributions as a function of $C$. The final result
is independent of $C$, which is a good way to check the computer code. 

Next we determine $F'(0)$ starting from eqn(\ref{eqn62}). Since the prefactor is
proportional to $x^2$, we must take the limit of $x\rightarrow 0$ with some
care. We use Ewald's method with an exponential 
\beq
\frac{1}{y_j} &=& \int_0^{\infty}dt e^{-ty_j}=\int_0^{\eta}dt
e^{-ty_j}+\int_{\eta}^{\infty}dt e^{-ty_j}\\
F_L'(x) &=&  -\frac{8x^2}{4\pi\sqrt{2}}\sum_{j\ne 0}\frac{e^{-
y_j(4x+\eta)}}{y_j}(1 +2xy_j + 2x^2y_j^2)\\
F_R'(x) &=&  -\frac{8x^2}{4\pi\sqrt{2}}\int_0^{\eta}dt\sum_{j\ne 0}e^{-
y_j(4x+t)}(1 +2xy_j + 2x^2y_j^2)
\eeq
The lattice sum vanishes at $x=0: F_L'(0)=0$. Evaluate $F_R'(x)$ by constructing
a periodic function of $\br$ and then determining its Fourier coefficient. 
\beq
V_R(\br) &=& \sum_{j}e^{- |\by_j-\br|(4x+t)}[1 +2x|\by_j-\br| +
2x^2(\by_j-\br)^2]=\sum_{\bcg}e^{i\bcg\cdot\br}v(G)\nonumber\\
v(G) &=& \frac{1}{\Omega_y}\int d^3r \exp[-i\bcg\cdot\br -r(4x+t)][1 +2xr +2
x^2r^2]\\
&=& 8\pi \sqrt{2}\left[\frac{\tau}{(\tau^2+G^2)^2} 
+2x\frac{3\tau^2-G^2}{(\tau^2+G^2)^3}+24x^2\tau\frac{\tau^2-G^2}{(\tau^2+G^2)^4}
\right]\\
\tau &=& 4x+t
\eeq
The result for $F'(0)$ comes from the term with $\bcg=0$. This integral is easy to evaluate 
\beq
v(0) &=&
8\pi\sqrt{2}\left[\frac{1}{\tau^3}+\frac{6x}{\tau^4}+\frac{24x^2}{\tau^5}
\right]\\
\int_0^{\eta}dt v(0) &=&
8\pi\sqrt{2}\left[\frac{1}{2}\left(\frac{1}{16x^2}-\frac{1}{(4x+\eta)^2}
\right)+2x 
\left(\frac{1}{(4x)^3}-\frac{1}{(4x+\eta)^3} \right)\right.\nonumber\\ &&
\left.+6x^2\left(\frac{1}{(4x)^4}-\frac{1}{(4x+\eta)^4} \right)\right]
\eeq
The terms in the above equation, that are proportional to $O(x^{-2})$, give the
limit as $x\rightarrow 0$
\beq
F'(0) &=& -\frac{8}{16}\left( 1 + 1 + \frac{3}{4}\right) =-\frac{11}{8}
\eeq
One can also use the Ewald sum to find the first nonzero term in $x$ which is
\beq
F'(x) &=& -\frac{11}{8} + B x^2 +O(x^3)\\
B &=& -\frac{\sqrt{2}}{\pi}\sum_{j\ne 0}\frac{e^{-
y_j\eta}}{y_j}+\frac{8}{\eta^2}+\frac{8\eta}{4\pi\sqrt{2}} -8\sum_{\bcg\ne
0}\frac{\eta^2}{G^2(G^2+\eta^2)}
\eeq
which gives $B = 1.4595 \pm 0.0001.$

\end{document}